\documentclass{svproc}

\usepackage{graphicx}        

\usepackage{url}
 
\begin{document}

\title{Greedy Navigational Cores in the Human Brain}
\titlerunning{Navigable Cores}  
%
\author{Zalán Heszberger\inst{1,2} \and András Majdán \inst{1,2} \and András Biró\inst{1} \and András Gulyás\inst{1,2} \and László Balázs\inst{1} \and Vilmos Németh\inst{3}
\and József Bíró \inst{1}}
\authorrunning{Heszberger et al.} 
%
%
\institute{Department of Telecommunications and Media Informatics \\ Budapest University of Technology and Economics, Budapest, Hungary \\
	\email{\{biro\}@tmit.bme.hu}
	\and
    MTA-BME Information Systems Research Group, Budapest, Hungary\\
    \and 
    BME Center of University-Industry Cooperation, Budapest, Hungary \\
    Regular Research Paper \\
    keywords: brain, navigation, greedy routing
	}

\maketitle    

\begin{abstract}

Greedy navigation/routing plays an important role in geometric routing of networks because of its locality and simplicity. This can operate in geometrically embedded networks in a distributed manner, distances are calculated based on coordinates of network nodes  for choosing the next hop in the routing. Based only on node coordinates in any metric space, the Greedy Navigational Core (GNC) can be identified as the minimum set of links between these nodes which provides 100\% greedy navigability. 
In this paper we perform results on structural greedy navigability as the level of presence of Greedy Navigational Cores in 
structural networks of the Human Brain.

\end{abstract}
\section{Introduction}
%
Greedy navigation is a type of hop-by-hop routing strategy in geometrically embedded networks. Geometric embedding means that nodes have either physical coordinates (e.g. 2D/3D Euclidean, or spherical) or inferred coordinates in a more abstract metric space (e.g. hyperbolic space).  The operation is as follows: Along the greedy route every node passes the information to that neighbouring node which is closest (and closer than the passing node) to the destination node in the metric space in which the network is embedded. Clearly, nodes have to know only the coordinates of their neighbours, based on these and the destination node coordinate (available typically with the information packet arrived) the distances can be calculated and the next hop as being closest to the destination can be identified. This is a simple operation using only local information providing distributed operation, that is why this type of routing is an actively researched area both in technological, natural and social  networks. 

Real networks having coordinates in a metric space are not necessarily 100\% greedy navigable. It means that greedy routing can get stuck in a node which has no direct link to the destination node and all of their neighbors are farther from the destination. Nevertheless, the level of greedy navigability (even if it is far from 100\%) may be an important sign of the co-evolution of the function-structure of the network and the efficiency/inefficiency of the metric space.   
For example, in \cite{natcomm2015} Gulyás et al. have tested several real networks against greedy navigability, and found that scale-free networks (Internet, US airport networks, metabolic networks, word networks) are highly navigable (80-90 \%) with 2D hyperbolic space, the non-scale free Hungarian road networks is well greedy navigable in the 2D Euclidean space ($84$\%). For this, our fundamentally new, so-called function-structure approach was to create first the minimalistic and 100\% greedy navigable skeleton, the so called greedy navigational core (GNC) network, and than to test the presence of these cores in real networks. The GNC is a result of solving heuristically the hard optimization problem of finding the minimum set of links providing maximum ($100$\%) greedy navigability. The GNC can also be obtained by the so-called network greedy navigation game. In this game nodes have the strategies to set up minimum number of links by which they can cover all non-neighbouring nodes by greedy next-hop forwarding. The Nash equilibrium of this game (previously referred to as Nash Equilibrium Network, NNG network) is the Greedy Navigational Core \cite{natcomm2015}. 
In the last couple of years thanks to the advancement in MRI-based imaging technologies brain structural networks have been widely investigated \cite{sporns2010networks,fornito2016fundamentals,sotiropoulos2019building}. For setting up these networks the brain cortex has been divided into parcels (ROI, region of interest) acting as nodes and (as a possible method) MRI-based Diffusion Spectrum Imaging (DSI) is used to explore the nerve fiber paths (travelling through the white matter) connecting the ROIs. The brain parcels have inherently 3D Euclidean coordinates which provides the possibility of testing the Euclidean greedy navigability of such networks.  We have shown first in \cite{natcomm2015} that the Greedy Navigational Core as minimal network that is maximally navigable by design presents substantially also in a five-subject based averaged structural brain network. More specifically it is found that the GNC precision (the ratio of the number of the GNC links included in the real network and the total number of GNC links, sometimes referred to as true positive rate) is $89$\% in this brain network. The GNC for this brain network was created by using only the 3D Euclidean coordinates of brain parcels, no other information was used. Subsequent studies have advocated  thoughts and results on greedy navigation of brain networks \cite{bassett2017small,Seguin2018,serrano2020,pappas2020structural,zhou2020efficient}. They highlighted that greedy navigation as a decentralized communication strategy is well suited to spatially embedded networks like brains. 
In \cite{Seguin2018,serrano2020} the authors have followed the well-established structure-function approach, namely  structural brain networks have been directly tested against the function greedy navigation in terms of different success measures \cite{muscoloni2019navigability}.   
The authors of \cite{pappas2020structural} have explicitly used our unorthodox function->structure method \cite{natcomm2015} to generate Greedy Navigational Cores (referred to as Nash Equilibrium Network, NNG network) and used them to predict resting-state functional connectivity with high accuracy.
In this study, using the function-structure approach we present detailed and elaborated results on structural greedy navigability of networks of human brain in five different scales, including the consistency, robustness and structural similarities.      


%
\section{Results}
%
\subsection{Individual Networks}
We have performed investigations on structural greedy navigability as the level of GNC precisions in 200 structural brain networks from 40 individual subjects at 5 different scales (these scales correspond to resolutions of 83, 129, 233, 463, 1015 nodes in the brain structural networks) \cite{betzel2016}. 
More details on the data set are in subsection Methods. For the GNC network generations only the physical (3D Euclidean) coordinates of the brain parcels were used, no other anatomical data or considerations were utilized.
First we present results on the original networks inferred, without any link removal (pruning). We found that the level of GNC precision (the ratio of GNC links being also in the brain network) is high and quite consistent (relatively low standard deviation around the mean) among the 40 brain networks within all scale, in spite of the fact that the 40 brain networks significantly differ from each other at all scales. The mean (and the standard deviation) of the GNC precisions (besides some other network parameters) in different scales (with increasing resolutions) are in Table~\ref{GNC precisions}.
\begin{table}
\centering
    \begin{tabular}{|c||l|l|l|l|l|}
    \hline
    & Scale1 & Scale2 & Scale3 & Scale4 & Scale5 
    \\
    \hline \hline
    \# Nodes & 83 & 129 & 233 & 463 & 1015 \\
    \hline
    Average \# Links in brain networks & $1119.4$ & $1975.5$ & $3799.27$ & $7246.48$ & $14254.8$ \\
    \hline
    Average \# Links in GNC & $177.7$ & $292.9$ & $553.4$ & $1153.3$ & $2656.6$ \\
    \hline
    Average Degree in brain networks & $26.97$ & $30.63$ & $32.61$ & $31.30$ & $28.09$ \\
    \hline
    Average Degree in GNC & $4.28$ & $4.54$ & $4.74$ & $4.98$ & $5.24$ \\
    \hline
    Mean of GNC Precision & $0.85$ & $0.88$ & $0.81$ & $0.70$ & $0.51$ \\
    \hline
    Standard Deviation of GNC Precision & $0.025$ & $0.019$ & $0.017 $ & $0.017$ & $0.025$
    \\ 
    \hline
    \end{tabular} \\ \hspace{4mm}
    \caption{Mean and Standard Deviation of GNC Precisions on different scales of brain structural networks}
    \label{GNC precisions}
\end{table}

\begin{figure}[hbt]
    \centering
    \includegraphics[width=.85\textwidth]{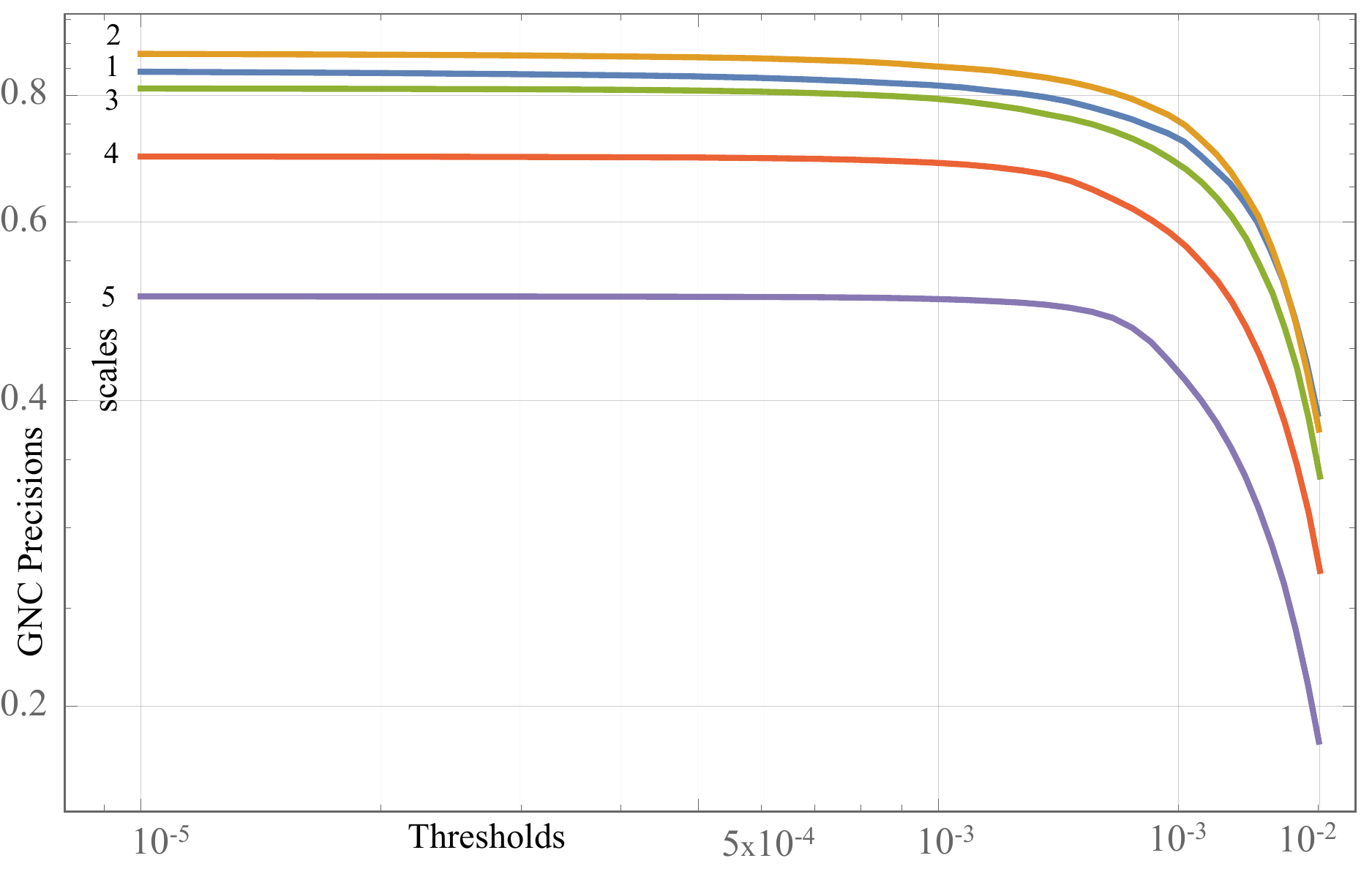}
    \caption{GNC Precision in the function of weight thresholds of pruning}
    \label{fig:GNC precision with pruning}
\end{figure}

{\bf Pruning by anatomical strength.} Now we turn to the case of network pruning, that is from the original networks links with low weights (possibly spurious nerve fiber paths) have been sequentially removed and these pruned networks have been tested against greedy navigability. For this, in every scale a sequence of networks has been generated with using different weight thresholds for pruning.
Table~\ref{GNC precisions pruned networks} summarizes the level of inclusion of GNC in the pruned networks for some characteristic threshold values.
For instance, when more than 50\% of the links are removed from the brain networks the precisions remain close to the original values.
One can also observe that larger part of the navigational core is missing from higher resolution brain networks, however, the precisions are still consistent according to the low standard deviations.
\begin{table}[]
    \centering
    \begin{tabular}{|c||c|c|c|c|c|c|}
    \hline
    Thresholds & \% of links removed &Scale1 & Scale2 & Scale3& Scale4 & Scale5   \\
    \hline \hline 
     $0$    & $0$\% &0.85(0.025) & 0.88(0.019) & 0.81(0.017) & 0.70(0.017) & 0.51(0.025) \\
     \hline
     $10^{-5}$ & $5-10$ \% & 0.84(0.024) & 0.88(0.019) & 0.81(0.017) & 0.69(0.017) & 0.51(0.025) \\
     \hline
     $10^{-4}$ & $23-29$ \% & 0.81(0.025) & 0.85(0.018) & 0.79(0.016) & 0.68(0.016) & 0.50(0.024) \\
     \hline
     $10^{-3}$ &$50-55 \%$ &0.72(0.031) & 0.75(0.019) & 0.68(0.020) & 0.57(0.019) & 0.42(0.021) \\
     \hline
     $10^{-2}$ & $82-84$ \% &0.39(0.034) & 0.37(0.033) & 0.34(0.028) & 0.27(0.022) & 0.18(0.017) \\
     \hline
    \end{tabular}
    \caption{GNC Precisions (Mean and Standard Deviation) in pruned networks. The first row shows the case without pruning.}
    \label{GNC precisions pruned networks}
\end{table}
A more detailed view of the effect of the network pruning can be seen in Fig.~\ref{fig:GNC precision with pruning}
One can observe that up to a certain threshold the GNC precision remains almost intact. Then further increasing the number of links cut out the GNC precision starts to decrease, however, even in case of extremely pruned networks it remains exceptionally high. This observations support the hypothesis that the GNC mostly consists of anatomically strong links, and spurious and/or anatomically weak connections are less likely part of the navigational cores.     

\begin{figure}
\centering
\includegraphics[width=.95\textwidth]{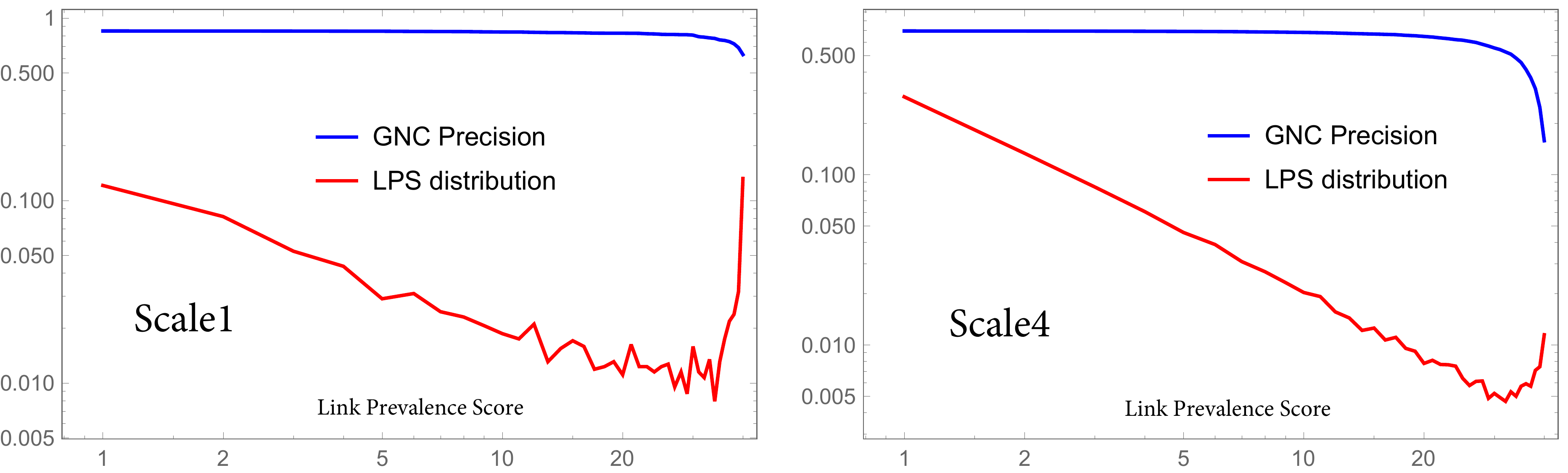}
\caption{Greedy Navigational Core precision and Link Prevalence Score distribution in the function of Link Prevalence Score}
\label{fig:GNC_vs_LPS}
\end{figure}

{\bf Pruning by Link Prevalence Score.} Besides anatomical weighting strategies, the link prevalence score (the number of networks containing the link) can also be used to identify possibly existent (high prevalence score) and possibly non-existent (low prevalence score) links in the inferred brain structural networks, and based on this one can compromise the false positives and false negatives in pruning the networks \cite{Heuvel2013}.
Here GNC precisions are also presented in \emph{individual} networks (the above mentioned 40-subject-5-scale networks with the number of nodes 83, 129, 234, 463, 1015 in scale 1, 2, 3, 4, 5, respectively) thresholded by the link prevalence scores. Thresholding means in this case that in an individual network only that links are kept, which are present at least $T-1$ other networks too, where $T$ is the threshold.  Within a resolution, every network is thresholded by all possible values of LPS's $(1,\ldots,40)$, then the GNC precisions are measured in all resulted networks (this corresponds sequence of 1600 networks in every scale). The GNC precisions are then averaged over the subjects for every LPS threshold. The important observation is that GNC precisions are still consistent (low variations across subjects) and robust against LPS thresholding in all scale. For example in scale 1 for LPS=1 (no link is removed) GNC precision is 0.85 and for LPS=30 (about 32\% of links are removed from every network) GNC precision is still as high as 0.80. In all the 5 scales it can be observed that for lower values of LPS thresholds the GNC precision remains almost intact while for higher values its decrease is fastening. The fastening decrease measurably coincides with the right hand side (consisting of possibly existing links) of the link prevalence distribution, see Fig~\ref{fig:GNC_vs_LPS}. This means that most of the true positive links in GNC networks are also possibly existent in the brain networks. 

\subsection{Average Networks}
In scale 5 we have constructed a sequence of average networks, based on averaging networks over link weights (inferred from measured anatomical strengths of fiber paths), and cutting out links with small average weights. We have also generated an average
GNC network based on averaging the centroids of the brain parcels. The size of the average GNC network (the number of links) is 2652.
Regarding the average network sequence, the following table shows the GNC precision in the function of the size of the average network:
\begin{center}
\begin{tabular}{|c||c|c|c|c|c|c|c|}
  \hline
  Threshold & $10^{-8}$ & $10^{-7}$ & $10^{-6}$ & $10^{-5}$ & $10^{-4}$ & $10^{-3}$ & $10^{-2}$ \\
  \hline
  Averaged Brain network size & 101468 & 101450 & 95929 & 67553 & 39406 & 15996 & 2381 \\
  \hline
  GNC Precision & 0.925 & 0.925 & 0.924 & 0.907 & 0.874 & 0.769 & 0.263 \\
  \hline
\end{tabular}
\end{center}

Note, that if none of the links have been cut out, the average brain network contains 101468 links (this corresponds to a $19.7\%$ connection density, because full mesh would have 514 605 links) and in this case 0.9253 fraction of the 2652 links of the
average GNC are in this network. If the number of links in the average network is quite comparable to the sizes of individual networks (Threshold=0.001, size of average
network=15996, $3\%$ connection density), the GNC precision is still as high as 76.9 \%. When the size of the average network is so extremely small that is comparable to the average GNC network, the GNC precision is still amazingly 26.28\%.

\subsection{Greedy Frames}

\begin{figure}[htb]
\centering
\includegraphics[width=1\textwidth]{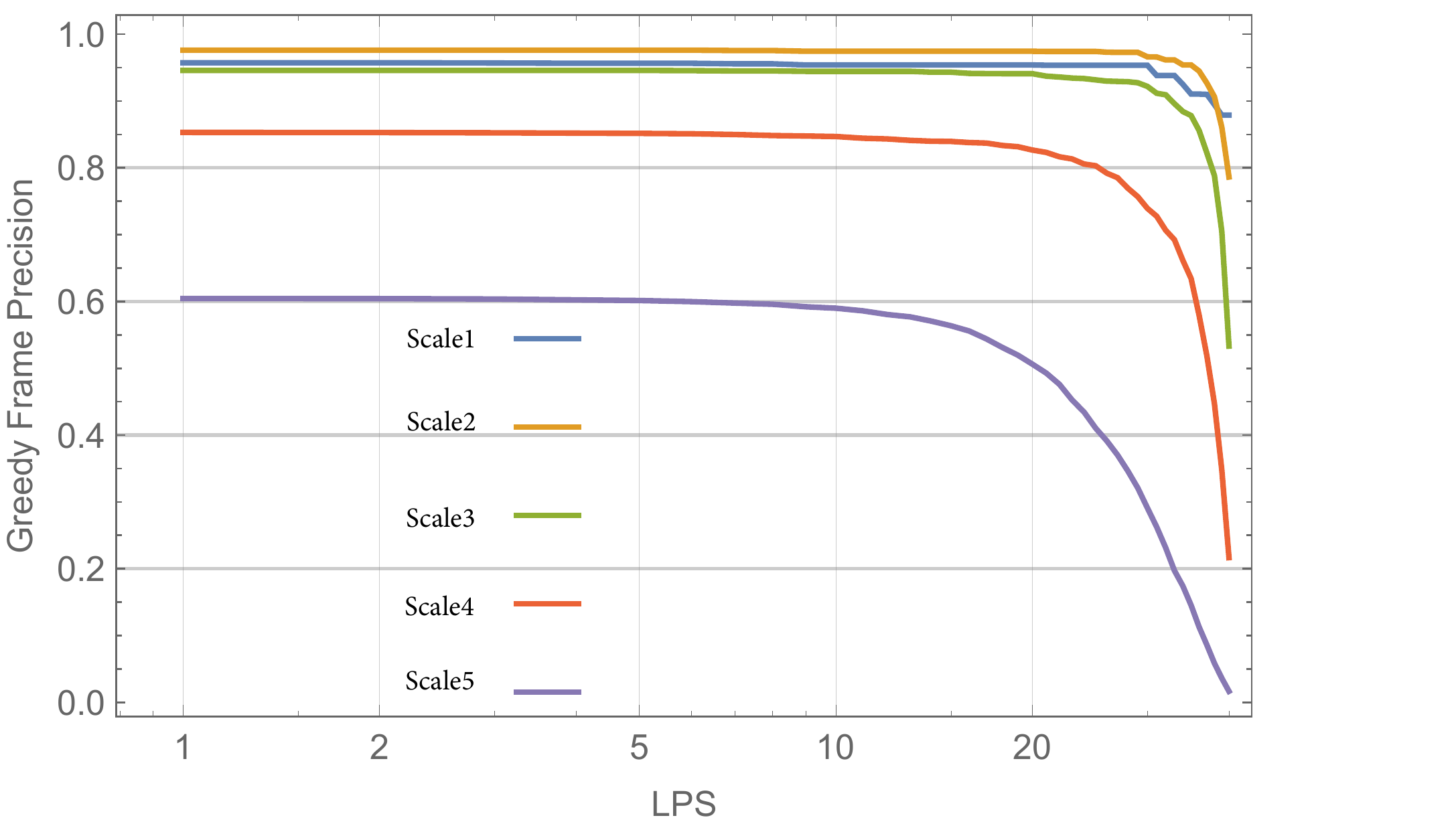}
\caption{Greedy Frame precision in the function of link pruning values based on Link Prevalence Score}
\label{fig:GFP_vs_LPS}
\end{figure}

The Greedy Navigational Core as the Nash equilibrium of the greedy network formation game is not unique. We always choose that one among these minimalistic networks which has the lowest aggregate link lengths. Nevertheless, there is always a common subset of the Nash equilibria called Greedy Frame \cite{natcomm2015}. These common links appearing in all maximally navigable minimal networks can be identified as follows. Let us take two nodes in the network, $u$ and $v$. If $v$ is the closest node to $u$ then for ensuring 100\% greedy navigability the (directed) link $v \rightarrow u$ must exist. Otherwise there would be no greedy path from $v$ (or through $v$) to $u$. The Greedy Frames can also be determined in our brain networks purely by using only the 3D coordinates of the parcels. The mean values (and standard deviation) of the Greedy Frame precisions (without pruning) in Scale 1 to Scale 5 are $ 0.942 (0.022), 0.985 (0.024), 0.947 (0.026), 0.832 (0.031), 0.595 (0.034)$ . One can observe that these inclusion ratios are even higher than the GNC precisions keeping the low variability within the scales. The inclusion ratio of the Greedy Frames are also robust against pruning. The Fig.~\ref{fig:GFP_vs_LPS} shows that cutting out links up to LPS=10 the Greedy Frame precisions do not change significantly in all scales. In case of lower resolutions (the first three scales) it is even true for LPS=30. In the highest resolution (scale 5) the Greedy Frame inclusion is remarkably lower than in other scales, but its decrease is more flat between LPS=20 and LPS=40.         

\section{Discussion}

The consistency and robustness of the greedy navigational core network precisions are remarkable in the light that these artificially generated networks are inferred only from the physical coordinates of the brain parcels by pure geometric computations and optimizations, no other anatomical data or consideration were used. 
The 40 brain networks within a scale show significant differences in terms of the links between the brain parcels. These differences originate from two main sources, measurement/imaging inaccuracy and anatomical variability. The effects of these can be decreased by pruning the networks (decreasing the false positive rates due to inaccurate imaging) and determining the optimal brain parcellation and parcel coordinates according to individual anatomy of subjects. As the results showed the GNC precision is significantly robust against pruning. This means that the common links of GNC and brain networks are likely existent and not spurious due to imaging artifacts. The greedy navigational cores can also trace the variations of brain networks through changes of the brain parcel coordinates, which results in keeping GNC precisions at high level with low variations.       







\begin{figure}[hbt]
\centering
\includegraphics[width=\textwidth]{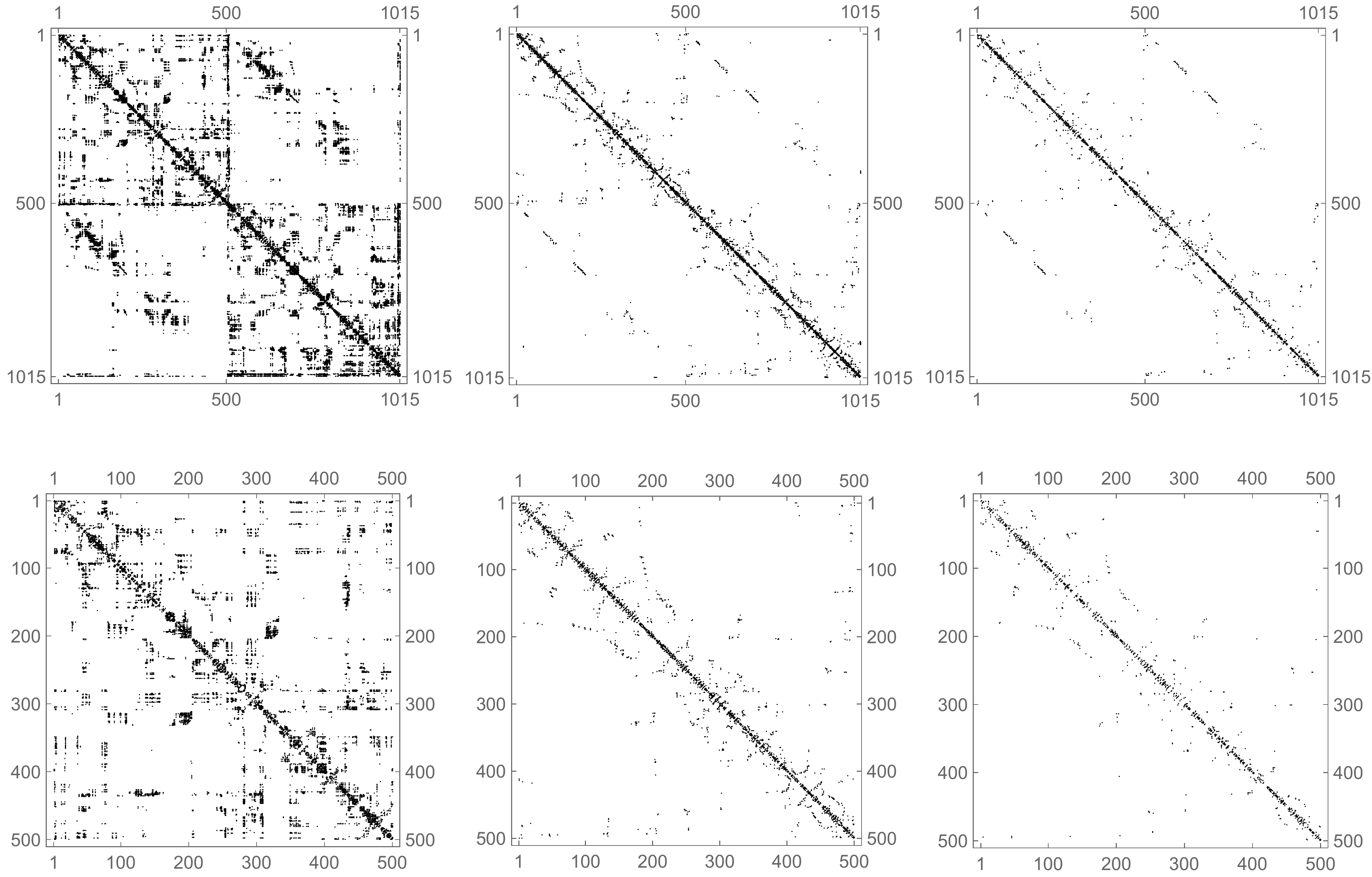}
\caption{Array plots of the adjacency matrix of a 1015 node (scale 5) structural brain network, the greedy navigational core (GNC) network, and that part of the GNC which is included in the brain network. Complete networks are in the first row, the upper-left part of the networks (500 nodes) are in the second row.}
\label{fig:matrixplots}
\end{figure}

GNC networks are not generative models for predicting brain links, because this is a minimalistic network containing much less links than the underlying brain network. This is also reflected by the average degrees, that is the GNC average degrees are between $4.2$ and $5.3$ while the brain network degrees lie between $26$ and $32$. In spite of this fact, the GNC networks show structural similarities to brain networks. This is illustrated by Fig.~\ref{fig:matrixplots} where one can observe that in a scale 5 GNC network and also in that part of the GNC which is included in the brain similar patterns can be identified as in the original brain network. Even the false positive links in GNC (that links which are not included in the brain) form such arrangements which seems to be a smooth "continuation" of the original brain patterns. This may relates to the fact that the false positive links in a GNC (corresponding to a given brain network) are likely present (with probability well above 0.9) in at least one of the remaining 39 brain networks. 

As the results showed the increased resolution of brain parcellation does not necessarily implies the decrease of GNC precision. Note that the highest inclusion ratios are in scale 2, the coarser scale 1 and finer scale 3 provide  somewhat lower precisions.  Significantly lower (but still high and consistent) precisions can be observed in scale 4 and 5. One can speculate that in these finer brain parcellations the 3D Euclidean space is not as suitable for GNC induction than in lower scales. One possible reason for this is that the brain cortex is highly folded and this may cause in higher resolutions that the lengths of curved fiber paths between brain parcels are less correlated to the Euclidean distances.  




\section{Methods}

{\bf Data.} The dataset used in our investigations contains 40 healthy human subjects who underwent an MRI measurement procedure where Diffusion Spectrum Imaging (DSI) data were obtained for each subject. The DSI data was processed according to the methods described in \cite{betzel2016}, resulting in 40 weighted, undirected structural connectivity maps comprising 83, 129, 233, 463, 1015 nodes in five different scales, respectively.
Each node represents a parcel of cortical or subcortical gray matter, and connections represent white matter streamlines connecting a pair of brain regions. Connection weights determine the average density of white matter streamlines and here consider connections with density above $10^{-8}$, resulting structural networks with an average of $1119$, $1976$, $3799$, $7246$ and $14254$ connections per subject.

\noindent {\bf Greedy Navigational Cores.} 
The greedy navigational core is generated from an empty network, using only the coordinates of nodes as input parameters. GNC can be considered as the solution of a  constrained optimization problem, in which the goal is to reach $100$\% greedy navigability with setting up minimum number of links between the nodes. This hard discrete optimization task can be traced back to the well-known minimum set cover problem, for solving that computationally efficient heuristics are available \cite{feige2004approximating}. 
Searching the GNC can also be formulated as a special network formation game (called network navigation game) in which the selfish players have strategies to set up links according to a payoff function in order to reach each others with greedy routes. An important property of GNC is that it is the Nash equilibrium of the network navigation game.

\section*{Acknowledgment}
Project no. 123957, 129589 and 124171 has been implemented with the support provided from the National Research, Development and Innovation Fund of Hungary, financed under the FK\_17, KH\_18 and K\_17 funding schemes respectively. Z. Heszberger and A. Gulyas have been supported by the Janos Bolyai Fellowship of the Hungarian Academy of Sciences and by the UNKP-19-4 New National Excellence Program of the Ministry of Human Capacities.

\bibliographystyle{unsrt} 
\bibliography{brain.bib}
\nocite{*}
\end{document}